\documentclass[prl,twocolumn,floatfix, showpacs]{revtex4}%

\usepackage{amsfonts}
\usepackage{amsmath}
\usepackage{amssymb}
\usepackage{graphicx}%

\begin{document}

\preprint{}

\title{Structure formation in electromagnetically driven  granular media}

\author{A.~Snezhko}
\author{I.S.~Aranson}
\author{W.-K.~Kwok}
\affiliation{Materials Science Division, Argonne National
Laboratory, 9700 South Cass Avenue, Argonne, IL 60439}



\begin{abstract}
We report structure formation in submonolayers of  magnetic
microparticles subjected to periodic electrostatic and magnetic
excitations. Depending on the excitation parameters, we observe
the formation of a rich variety of structures: clusters, rings, chains,
and networks. The growth dynamics and shapes of the structures are
strongly dependent on the amplitude and frequency of the external
magnetic field. We find that for pure ac magnetic driving at
low densities of particles, the low-frequency magnetic excitation
favors clusters while high frequency excitation favors chains and net-like
structures. An abrupt phase transition from chains to a network phase was observed for a high density of particles.
\end{abstract}


\received[Received: ]{20 October, 2004}%

\maketitle

Large  assemblies of macroscopic particles subject to external
driving, such as vibration, shear or rotation, exhibit fascinating
collective behaviors.  Their dynamics is poorly understood,
especially when inter-particle interactions are strongly
dissipative \cite{kadanoff}. Furthermore, additional complications
arise when the grain size goes below $0.1$ mm, and non-trivial
interactions due to charging or magnetization come into play. When
small particles acquire an electric charge or magnetic momentum,
the dynamics are governed by the interplay between long-range
electromagnetic and short-range contact forces. On the other hand,
precise electromagnetic excitation can be used to control the
morphology of the resulting pattern.

Recently, a number of experimental studies was performed with
vibrofluidized magnetic particles \cite{blair,stambaugh}. Several
interesting phase transitions were reported, in particular the
formation of dense two-dimensional clusters and loose
quasi-onedimensional chains and rings. Besides direct interest to
the physics of granular media, these studies may provide insight
into the fundamental problem of dipolar hard sphere fluids where
the nature of solid/liquid  transitions is still debated
\cite{pinkus}. Alternatively, vibrofluidized magnetic particles
can be considered as an extremely simplified model of a
ferrofluid, where similar experiments are technically difficult to
perform. Previous studies were limited to very small number of
particles (about 1,000) due to the intrinsic limitation of the
mechanical vibrofluidization technique.

We developed a technique to electrostatically drive fine
conducting powders \cite{aranson1,aranson2}. Our approach offers
unique new opportunities compared to traditional vibration
techniques. It enables one to deal with extremely fine powders
which are not easily controlled by mechanical methods. In addition
electrostatic driving allows to control the ratio between
long-range electric forces and short-range collisions by changing
the amplitude and the frequency of the applied electric field.

In this Letter, we extend our studies towards magnetic
microparticles. In addition to electrostatic excitation, we
explore the driving of magnetic particles using an ac magnetic
field. Our studies reveal a rich diversity of behaviors, including
formation of dense immobile clusters, as well as
quasi-onedimensional chains and rings. We provide strong evidence
of a first order phase transition from finite length chains of
particles to infinite networks as the driving parameters are
varied.

\begin{figure}[ptb]
\includegraphics[width=8.5cm]{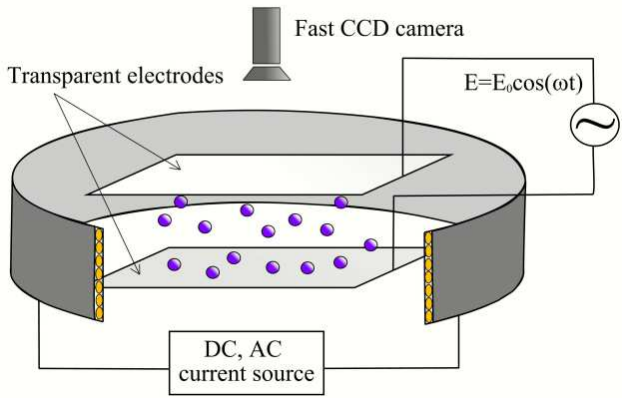}%
\caption{Schematic view of the experimental setup.}%
\label{fig1}%
\end{figure}

{\it The experimental setup} is shown  in Fig.~\ref{fig1}. The
design of the electrostatically driven cell is similar to that
reported earlier ~\cite{aranson1}. Conductive particles are placed
between two horizontal transparent conducting glass plates
(12$\times$12 cm  with the spacing of 1.5 mm). An external
magnetic field is provided by a set of large magnetic coils (30 cm
in diameter) placed around the cell. To excite the granular media
contained in the cell, a voltage of 0 - 2 kV with a frequency of 0
- 150 Hz is applied to the plates. The magnetic interaction is
controlled by supplying a dc
 electric current to one of the coils to create a constant
magnetic field in the range of 0-80 Oe. In addition, pure
"magnetic" driving of the particles is carried out by applying an
ac current to a second coil with small or no electric field
applied to the cell. The amplitude of the ac magnetic field can be
varied from 0-15 Oe with the frequency from 0 to 300 Hz. The
granular media used in our experiment consisted of spherical
Nickel microparticles with average size of about 90 $\mu$m (Alfa
Aesar Company). Magnetic moment per particle at the 80 Oe field is
$~1\cdot10^{-5}$ emu;   saturated magnetic moment is
$2\cdot10^{-4}$ emu per particle,  saturation field is about 4kG.
The number of particles  was varied in the range $120,000 -
300,000$. Experiments were also performed with 40 $\mu$ sized
particles, but no qualitative difference was found. The
experiments were carried out in air and in non-polar low-viscosity
liquid (toluene). Real time images were acquired with a
high-speed- digital camera.

Electrostatic driving  works as follows. Conducting particles
acquire an electric charge when they are in contact with the
bottom conducting plate of the cell. When the magnitude of the
electric field in the cell exceeds some critical value, the upward
electric force acting on the charged particle overcomes the
gravitational force and the particles are driven towards the upper
plate. Upon contact with the upper plate, the particles recharge
and fall back to the bottom plate. This process repeats in a
cyclical fashion.  The elevation of the particles off the bottom
plate can be adjusted by the frequency of the applied ac electric
field. Pure "magnetic" driving works in the following way.
Magnetic particles with moment \emph{\textbf{M}} subjected to an
external magnetic field \emph{\textbf{H}} experience a torque
$-[\emph{\textbf{M}}\times \emph{\textbf{H}}]$ forcing the
magnetic moment of particles to be aligned with the applied
magnetic field. There are, however, two ways in which the magnetic
moment of the particle can adjust its direction: (i) the magnetic
moment can rotate inside the particle against the internal
magnetic anisotropy field and (ii)the whole particle can rotate in
order to keep the moment aligned with the applied magnetic field.
In the latter case, the magnetic particle needs to overcome the
resistance from the friction and adhesion forces between the
particle and surface of the plate.
Thus, as the magnitude of the external magnetic field exceeds some
critical value, each particle begins to rotate by keeping its
moment aligned with the field and moves being driven by magnetic
drag force
$\emph{\textbf{F}}_{m}=\textbf{M}\nabla\textbf{H}_{local}$ where
$\textbf{H}_{local}$ designates the local magnetic field coming
from dipolar fields of the neighboring particles and the external
magnetic field. The effective time the particle can move may be
controlled by the adjusting the frequency of an ac magnetic field.

\begin{figure}[tb]
\includegraphics[width=8.5cm]{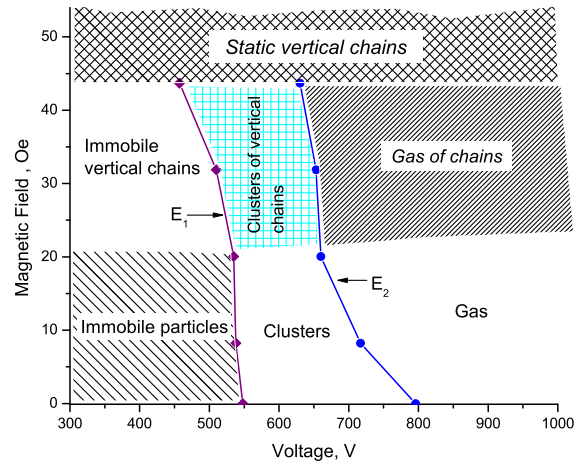}%
\caption{Phase diagram in the magnetic field, voltage plane for $90\mu$m Nickel particles in the air filled cell. Surface fraction
of particles is 9\% ($\approx$ 200,000 particles)}%
\label{fig2}%
\end{figure}

As reported earlier, ~\cite{aranson1,aranson2} the particles
remain immobile on the bottom plate of the cell if the applied
electric field is less than some critical value $E_{1}$. When the
field exceeds a second threshold value, $E_{2} > E_{1}$, the
system undergoes a transformation to a gas-like state. When
decreasing the electric field $E$ below $E_{2}$ ($E_{1}<E<E_{2})$,
small densely packed clusters begin to nucleate. Clustering is
promoted through electrostatic screening (two  particles in
contact acquire a smaller charge than two well-separated
particles). In the case of magnetic particles, an additional
complication arises from the dipole-dipole magnetic interactions
and the dynamics of cluster formation and critical electric fields
$E_{1}, E_{2}$ are expected to change.  The phase diagram
delineating the primary experimental regimes as a function of
applied voltage and external magnetic field
 for  200,000 nickel particles with average size of 90 $\mu$m  (about 9\% monolayer
coverage) is shown in Fig.~\ref{fig2}. The frequency $f$ and
amplitude $V$  of the ac electric field applied to the cell were
chosen to provide quasi-two dimensional motion of the particles
(V$<$1000V, $f=100$Hz). The cell was subjected to an external dc
magnetic field from 0 to 80 Oe.

\begin{figure}[b]
\includegraphics[width=8.5cm]{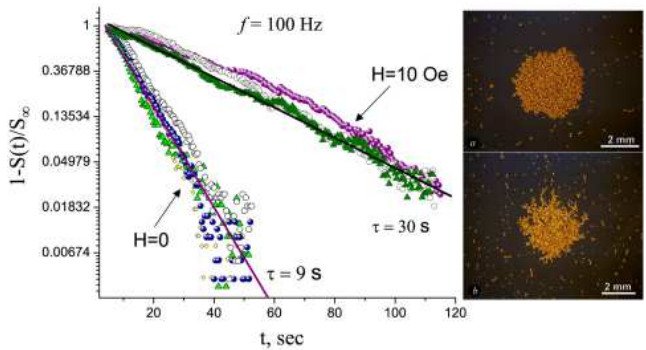}%
\caption{Rescaled cluster area vs time at zero (a) and 10 Oe (b)
magnetic fields. Solid lines represent simple functional
dependence (\ref{1})  for different characteristic growth times $\tau$.}%
\label{fig3}%
\end{figure}

The behavior of the magnetic particles in low external magnetic
fields (below 20 Oe) is rather similar to that reported for
nonmagnetic conductive particles. The isolated particles are
immobile until the electric field exceeds a critical value
$E_{1}$. At  electric field values above  $E_{2}$, the granular
medium transforms into a uniform gas-like phase. Cluster formation
is observed in the interval, $E_{1}<E<E_{2}$, in agreement with
earlier studies on nonmagnetic granular systems
~\cite{aranson1,sapozhnikov}. However, the clustering dynamics of
the magnetic particles is very sensitive to the external magnetic
field. Clusters formed in the absence of a magnetic field (a) and
at a magnetic field of 10 Oe (b) are shown in Fig.~\ref{fig3}. The
cluster grown in an external magnetic field looks less compact and
sufficiently less smooth in shape. Moreover, cluster formation in
an external magnetic field is characterized by arrested coarsening
due to depletion of the gas phase.

We analyzed the growth rates to characterize the time evolution of the clusters  at different
magnetic fields. We find that the change of the cluster size for both zero and
nonzero external magnetic fields can be well described by a simple
equation:
\begin{equation}
S(t)=S_{\infty}\left( 1-\exp\left( -(t-t_{0})/\tau
\right)\right)\label{1}
\end{equation}
Here $S_{\infty}$ is the cluster size at $t\to \infty$; $t_{0}$ is
the time when the cluster nucleates; $\tau$ is the growth time.
The quantity $(1-S(t)/S_{\infty})$ for the clusters formed at zero
field and at 10 Oe is plotted in Fig.~\ref{fig3}. Solid lines in
Fig.~\ref{fig3} represent the functional dependence of Eq.
(\ref{1}) with the corresponding characteristic growth times.  The
larger growth time $\tau$ in the presence of an external magnetic
field could be explained as follows: since the particles are large
enough to contain multi-domains, the higher external magnetic
field induces larger magnetic moments in the particles (due to the
growth of magnetic domains within the particle with orientation
along the external field) leading to amplification of the
dipole-dipole interactions. Consequently, the interaction between
particles becomes highly anisotropic and increases the
characteristic growth time since (a) aggregation proceeds
predominantly through coalescence of chain segments which are less
mobile; (b)  some particles are repelled from the cluster due to
unfavorable magnetic moment orientation. For elevated magnetic
fields (20-45 Oe) the formation of immobile vertical chains
consisting of two to four particles were observed when the
amplitude of the electric fields was reduced below $E_{1}$. The
threshold electric field $E_{1}$ decreases with increasing
external magnetic field since the formation of immobile chains
diminishes the effective spacing between cell plates resulting in
higher electric field intensity for particles at the top parts of
the chains. Upon exceeding $E_{1}$ chains start to move and form
localized clusters of chains. At amplitudes of electric field
above $E_{2}$, the system transforms to a gas of short chains.
Increase of the magnetic field above 50 Oe creates static long
vertical chains touching the upper plate of the cell.
\begin{figure}[t]
\includegraphics[width=8.5cm]{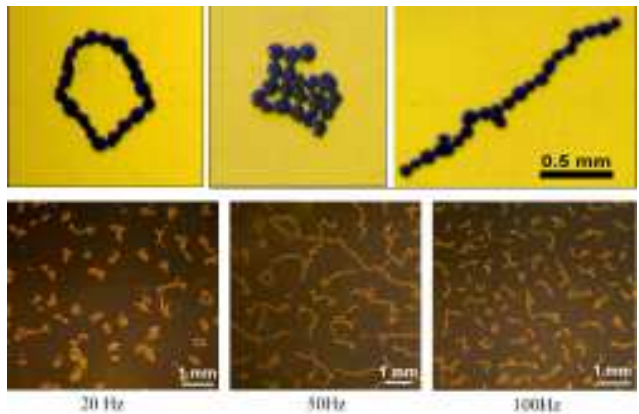}%
\caption{Structures formed  in external \emph{ac} magnetic field:
 ring; compact clusters;  chains of dipoles. \emph{Bottom:} Patterns
formed by nickel spheres (0.053 of the surface monolayer coverage)
under magnetic driving at 20 Hz (clustered phase), 50 Hz (net-like
structure), and 100 Hz (chains phase).}%
\label{fig4}%
\end{figure}

We explored pure ac magnetic driving by placing the magnetic
particles into a cell filled with toluene to dampen the  kinetic
energy in the system. Before each experiment, the system was
driven to a gas-like state with an ac electric field in zero
magnetic field to create a uniform distribution of particles over
the cell. Subsequently, the electric field was turned off and the
system was subjected to \emph{ac} $15$ Oe magnetic field with
frequency from 0 to 200 Hz.  The magnetic dipole-dipole
interaction favors completely different particle organization.
Fig.~\ref{fig4}\emph{(upper part)} demonstrates some selected
structures.

The nature of the self-assembled structures is quite
straightforward. Since the local arrangement is dominated by a
highly anisotropic dipole-dipole magnetic interaction, the
head-to-tail interaction is the strongest, favoring formation of
the quasi-one dimensional chain structures. Each chain has a
certain rigidity against bending since in a bent chain the dipoles
are not fully aligned. When the bending angle exceeds some
threshold value, the energy barrier to bending starts to decrease
since closing of the chain now becomes more favorable due to the
attraction of the opposite ends of the chain. It can be shown that
a ring is more energetically favorable than a chain if the number
of particles in the chain exceeds four ~\cite{wen}. However, to
form a ring from a chain, one needs to overcome a strong potential
barrier associated with chain bending. Consequently, a more
plausible mechanism for forming a ring is a "fusion" of several
short chain segments of appropriate orientation, polarization and
position, see \cite{aux}. Chains can also form compact structure,
such as a cluster of dipoles. Such dipole clusters can nucleate in
the situation when two neighboring parallel chains have opposite
polarization and attract each other to form a cluster.

One can vary the frequency of the ac magnetic field to control the
time it takes the particles (or chains) to rotate in order to keep
its moment aligned with the local field and to move along the
local field gradient. If the frequency is too high (200Hz in our
case), nothing happens since the characteristic reaction time of
the system is higher than the period of the ac magnetic field. At
some point while decreasing the frequency, the particles start to
react to the external ac magnetic field and short chains appear in
the system. A further decrease of the frequency leads to the
creation of more energetically favorable configurations such as
rings and branched chains.

\begin{figure}[ptb]
\includegraphics[width=8.3cm]{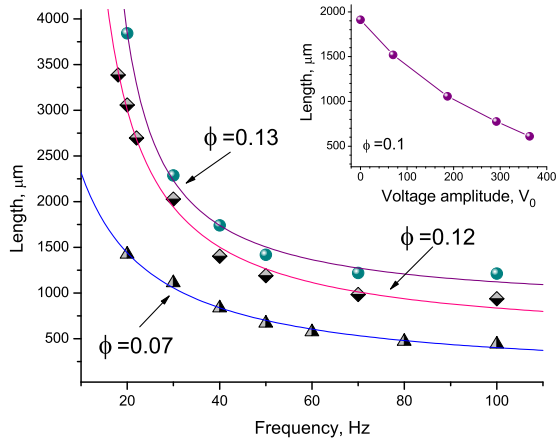}%
\caption{Saturated chain length vs frequency of  applied 15 Oe
\emph{ac} magnetic field  for different amount of nickel $90 \mu
m$ particles in the cell. Solid lines are the fits to expression
$y=A+B/(x-f_{0})$. \emph{Inset:} Saturated chain length vs applied
ac (100Hz) electric voltage amplitude for magnetically driven
($H_{max}=15$ Oe; f=25 Hz) system ($\phi\simeq$0.10).
 }%
\label{fig5}%
\end{figure}
Compact clusters as shown in  Fig.~\ref{fig4} can be initiated by
decreasing the frequency of the ac magnetic field. The bottom
panel of Fig.~\ref{fig4} shows various patterns formed under the
influence of an external ac magnetic field at 20, 50, and 100 Hz
for a cell with 5.3\% surface coverage of Ni particles. Each
experiment was started from randomly dispersed configuration of
the nickel spheres over the bottom plate of the cell. Clearly, the
pattern is determined by the frequency of the external ac magnetic
field: low-frequency excitations (0-30 Hz) produce a clustered
phase; high-frequencies (80-200 Hz) assemble short chains while
intermediate-frequencies favors net-like patterns. The resulting
pattern is strongly history dependent. For example, a continuous
decrease of the magnetic field frequency from 200 to 10 Hz results
in the formation of a net-like structure, in contrast to the
formation of clusters obtained when the system "relaxes" at a
constant frequency of 10 Hz. Denser configurations ($\phi$=0.115
and 0.133), however, do not exhibit the cluster phase and instead,
a phase transition to the network phase (infinitely long
multi-branch chains) is observed at low frequencies. To
demonstrate the phase transition to the network phase, the average
chain length as a function of the \emph{ac} magnetic field
frequency is plotted in Fig. ~\ref{fig5} for different densities.
Data was taken after the system was left to "relax" in the applied
field for about 10 minutes to attain their respective equilibrium
state. As the frequency of the magnetic field decreases, the
average length of the chains tends to diverge for dense
configurations ($\phi$=0.115 and 0.133) and saturates for less
dense ones. The solid lines in the figure are fits to the
expression, $A+B/(x-f_{0})$, where $f_{0}$ designates a critical
frequency when the system undergoes a phase transition to the
net-like phase. The critical frequencies extracted from the fits
are 14.10 Hz, and 6.11 Hz for $\phi$=0.133, and $\phi$=0.115,
respectively. The critical frequency for surface coverage of
$\phi$=0.07 resulted in a negative value, $f_{0}$=-5.62 Hz,
indicating that there is no transition to the network phase.
Indeed, for such low density configurations, compact clusters were
observed at low frequencies.

    Inset to Fig.~\ref{fig5}
demonstrates the effect of an \emph{ac} electric field on  the
system of 240000 Ni particles ($\phi\simeq$ 0.10) in  ac magnetic
field ($H_{max}=$15 Oe; f=25 Hz). The frequency of the electric
field was kept at 100 Hz to provide mostly two-dimensional motion.
The average length of the chain decreases with increasing
\emph{ac} electric field, suggesting that the latter acts as an
analogue to temperature.

We studied the self-assembly of magnetic microparticles in ac
electric  and  magnetic field.
Excitation of the system by an ac magnetic field revealed a
variety of patterns that can be controlled by adjusting the
frequency and the amplitude  of the  field. We found that at low
particle densities the low-frequency magnetic excitation favors
cluster phase formation while high frequency excitation favors
chains and net-like structures.  For denser configurations  an
abrupt transition to the network phase was observed. We thank
Ulrich Welp for help with  magnetization measurements. This
research was supported by the US DOE, grant W-31-109-ENG-38.

\end{document}